\RequirePackage{fix-cm}
\RequirePackage{lineno}
\documentclass[smallcondensed]{svjour3}     
\smartqed  
\usepackage[colorlinks=true, citecolor={blue}, urlcolor={black}]{hyperref}
\usepackage{graphicx}
\usepackage{amsmath}
\usepackage{amssymb}
\usepackage{color}
\usepackage{longtable}
\usepackage{afterpage}
\usepackage{caption}
\usepackage{latexsym}
\usepackage{array}
\usepackage{lineno}
\newcolumntype{L}[1]{>{\raggedright\let\newline\\\arraybackslash\hspace{0pt}}m{#1}}
\newcolumntype{C}[1]{>{\centering\let\newline\\\arraybackslash\hspace{0pt}}m{#1}}
\newcolumntype{R}[1]{>{\raggedleft\let\newline\\\arraybackslash\hspace{0pt}}m{#1}}

\begin{document}

\title{ Energy sensitivity of the GRAPES-3 EAS array for primary cosmic ray
protons \thanks{$^\ast$Corresponding author}}

\author{
        B.~Hariharan        \and
	    S.~Ahmad            \and	
	    M.~Chakraborty      \and
	    A.~Chandra	        \and
        S.R.~Dugad          \and
        S.K.~Gupta          \and
        Y.~Hayashi          \and
        H.~Kojima           \and
        S.S.R.~Inbanathan   \and
        P.~Jagadeesan       \and
        A.~Jain             \and
        P.~Jain		        \and
        V.B.~Jhansi         \and
        S.~Kawakami         \and
        P.K.~Mohanty        \and
        S.D.~Morris         \and
        P.K.~Nayak          \and
        A.~Oshima           \and
        D.~Pattanaik        \and
        P.S.~Rakshe         \and
        K.~Ramesh           \and        
        B.S.~Rao            \and
        L.V.~Reddy          \and
        S.~Shibata          \and
        F.~Varsi            \and
        M.~Zuberi
}

\institute{
    B.~Hariharan$^\ast$ \and S.~Ahmad \and M.~Chakraborty \and A.~Chandra \and
    S.R.~Dugad \and S.K.~Gupta \and Y.~Hayashi \and H.~Kojima \and P.~Jagadeesan
    \and A.~Jain \and P.~Jain \and V.B.~Jhansi \and S.~Kawakami \and
    P.K.~Mohanty \and S.D.~Morris \and P.K.~Nayak \and A.~Oshima \and
    D.~Pattanaik \and P.S.~Rakshe \and K.~Ramesh \and B.S.~Rao \and L.V.~Reddy
    \and S.~Shibata \and F.~Varsi \and M.~Zuberi \at 
            The GRAPES-3 Experiment, Cosmic Ray Laboratory, Raj Bhavan, Ooty
            643001, India \\
            \email{89hariharan@gmail.com} 
            \and
    B.~Hariharan$^\ast$ \and M.~Chakraborty \and S.R.~Dugad \and S.K.~Gupta \and 
    P.~Jagadeesan \and A.~Jain \and V.B.~Jhansi \and 
    P.K.~Mohanty \and S.D.~Morris \and P.K.~Nayak \and D.~Pattanaik \and 
    P.S.~Rakshe \and K.~Ramesh \and B.S.~Rao \and L.V.~Reddy \at
            Tata Institute of Fundamental Research, Dr Homi Bhabha Road, Mumbai
            400005, India \\              
            \and
    B.~Hariharan$^\ast$ \and S.S.R.~Inbanathan$^\ast$ \at
            Post Graduate \& Research Department of Physics, 
            The American College, Madurai 625002, India \\
            \email{89hariharan@gmail.com, ssrinbanathan@gmail.com}
            \and
    S.S.R.~Inbanathan$^\ast$ \at
            Department of Applied Science, 
            The American College, Madurai 625002, India \\
            \email{ssrinbanathan@gmail.com}
            \and
    S.~Ahmad \and A.~Chandra \and M.~Zuberi \at
            Aligarh Muslim University, Aligarh 202002, India \\
            \and
    Y.~Hayashi \and S.~Kawakami \at
            Graduate School of Science, Osaka City University, 558-8585 Osaka,
            Japan
            \and
    H.~Kojima \and A.~Oshima \and S.~Shibata \at
            College of Engineering, Chubu University, Kasugai, 
            Aichi 487-8501, Japan
           \and
    P.~Jain \and F.~Varsi \at
            Indian Institute of Technology Kanpur, 
            Kanpur 208016, India
}

\date{Received: date / Accepted: date}
\maketitle

\begin{abstract}
Low energy ground-based cosmic ray air shower experiments generally have energy
threshold in the range of a few tens to a few hundreds of TeV. The shower
observables are measured indirectly with an array of detectors. The atmospheric
absorption of low energy secondaries limits their detection frequencies at the
Earth's surface. However, due to selection effects, a tiny fraction of low
energy showers, which are produced in the lower atmosphere can reach the
observational level. But, due to less information of shower observables, the
reconstruction of these showers are arduous. Hence, it is believed that direct
measurements by experiments aboard on satellites and balloon flights are more
reliable at low energies. Despite having very small efficiency ($\sim$0.1\%)
at low energies, the large acceptance ($\sim$5\,m$^2$sr) of GRAPES-3 experiment
allows observing primary cosmic rays down below to $\sim$1\,TeV and opens up
the possibility to measure primary energy spectrum spanning from a few TeV to
beyond cosmic ray knee (up to 10$^{16}$\,eV), covering five orders of magnitude.
The GRAPES-3 energy threshold for primary protons through Monte Carlo
simulations are calculated, which gives reasonably good agreement with data.
Furthermore, the total efficiencies and acceptance are also calculated for
protons primaries.  The ability of GRAPES-3 experiment to cover such a broader
energy range may provide a unique handle to bridge the energy spectrum between
direct measurements at low energies and indirect measurements at ultra-high
energies.
\end{abstract}

\section{Introduction}

Energy spectrum and composition studies are the key objectives of any cosmic ray
(CR) experiment to understand its origin and acceleration mechanisms.  Majority
of the CRs are lighter elements such as proton (90\%), helium (9\%), and the
remaining (1\%) are heavier elements including carbon, nitrogen, oxygen,
aluminum, and iron. The primary cosmic rays (PCRs) have been observed over an
extraordinary energy range of 10$^8$--10$^{20}$\,eV, spanning twelve orders of
magnitude. The energy spectrum is represented by power law distribution with two
prominent features including the ``{\em knee}" at about
$\sim$3$\times$10$^{15}$\,eV where the spectral index changes from -2.7 to -3.1
and ``{\em ankle}" at $\sim$3$\times$10$^{18}$\,eV where the spectral index
again changes from -3.1 to -2.8. It is believed that these features are due to
transition of CR sources from galactic to extragalactic origin. However, this
still remains an unsolved mystery. 

Upon entering into Earth's atmosphere, the PCR collides with atmospheric gaseous
molecules and produces secondary particles that includes pions, kaons, etc. The
charged pion decays into a muon and its associated neutrino where the muon
subsequently decays into an electron and two neutrinos. The neutral
pion decays into two $\gamma$-rays that further develop into an electromagnetic
cascade through bremsstrahlung and pair production processes. These chain of
interactions are continued until the particles are decayed or stopped as they
propagate down to observational level. The entire process of the development is
called extensive air shower (EAS) or cascade shower. At a particular stage, the
shower development reaches its maximum number of particles, which is known as
shower maximum and it can be represented by $X_{max}$. The shower development is
also characterized by a parameter called age ($s$) where $s$=0 corresponds to
the first interaction, $s$=1 at the shower maximum, and $s$=2 represents the
death of the shower. The number of secondaries and their lateral spread in an
EAS at the observational level depends on PCR energies. For an ultra-high energy
primary ($>$10$^{18}$\,eV), the number of particles may reach in excess of
billions spread over hundreds of square kilometers. The EAS measurement
technique is used to determine the information by sampling the secondaries at
the observational level in the particle detectors. Typical air shower
experiments consist of an array of detectors deployed over a large area. The
primary parameters are reconstructed by using the information from the
secondaries with aid of Monte Carlo simulations. 

Direct measurements of PCRs are carried out by detectors aboard on balloons or
satellites. However, these instruments do not have enough sensitivity above
$\sim$100\,TeV due to low flux of PCRs, short exposure time, and limited
detector size. Satellite based experiments such as PAMELA \cite{pamela_1},
AMS-02 \cite{ams2_1}, and DAMPE \cite{dampe} and balloon-borne experiments such
as CAPRICE \cite{caprice}, BESS \cite{bess}, and CREAM \cite{cream_1} primarily
use calorimeters for energy measurements and particle's time of flight (TOF) in
the detector for measuring arrival directions of the primaries. The CRs energy,
composition, and arrival direction measurements done by balloon and satellite
based experiments are fairly precise. However, they lack statistics at higher
energies, which restrict their measurements to below $\sim$100\,TeV. The PCRs of
energy above $\sim$100\,TeV are indirectly studied by using EAS technique with
an array of detectors placed on the ground level like ARGO-YBJ \cite{argo},
Tibet AS$\gamma$ \cite{tibetas} KASCADE \cite{kascade}, KASCADE-Grande
\cite{grande}, AUGER \cite{auger}, and TA \cite{ta}. However, these experiments
are sensitive at different energy ranges such that ARGO-YBJ and Tibet AS$\gamma$
which have operating energy from few a TeV to below the knee region
($\sim$10$^{15}$\,eV), KASCADE and KASCADE-Grande operated around knee region,
and experiments like AUGER and TA are mainly aimed to explore the origin of
ultra high energy cosmic rays so they are sensitive to much higher energies
beyond knee ($\sim$10$^{19}$\,eV). 

The energy threshold of an EAS experiment is decided by the inter-detector
separation whereas its total physical area covered by decides its upper energy
end. Generally, most of the low energy showers get absorbed in the atmosphere
and do not reach up to ground level. However, a small fraction of young low
energy showers, which are produced deep into the atmosphere may survive till
ground level. However, these showers have very low triggering efficiencies and
recorded with limited information, which makes their reconstruction difficult.
Due to these low efficiencies, the ground-based experiments generally make
observations at higher energies. So, historically it is believed that the direct
measurements are more reliable at lower energies (below $\sim$100\,TeV) whereas
indirect measurements stay above these energies.  In such scenario, the
contributions of experiments from ARGO-YBJ, Tibet AS$\gamma$, and GRAPES-3 are
vital if the low energy PCRs can be used to provide in the extended energy range
of TeV--PeV to bridge direct and indirect measurements. The energy threshold
of proton initiated EAS for GRAPES-3 is estimated to be few TeV by using Monte
Carlo simulations with reasonably good agreement with data. From these
simulations, the triggering and reconstruction efficiencies, and geometrical
acceptance are also estimated as a function of energy.  It is found that only a
small fraction (i.e.  $\sim$0.1\% of incident) of such low energy showers are
detected and reconstructed, which can be used for further studies.  Due to the
larger physical area coverage, the geometrical acceptance of GRAPES-3 is larger
compared to direct measurements like PAMELA (20.5\,cm$^2$sr) \cite{pamela_2},
AMS-02 (0.5\,m$^2$sr) \cite{ams2_2}, DAMPE (0.1\,m$^2$sr) \cite{dampe}, BESS
(0.3\,m$^2$sr) \cite{bess}, and CREAM (0.322\,m$^2$sr) \cite{cream_2}. The
GRAPES-3 acceptance is estimated to be $\sim$5\,m$^2$sr for 1\,TeV protons.
Maximum efficiency has been achieved at $>$100\,TeV with the acceptance of
$>$20000\,m$^2$sr.  Though, the efficiencies are small at low energies but due
to immense flux of incident primaries, and larger acceptance, the GRAPES-3
records large number of usable low energy EAS that can provide energy spectrum
and mass composition of PCRs in the broad energy span to overlap with low and
ultra-high energy measurements from direct and indirect experiments
respectively.

\section{The GRAPES-3 experiment}

\begin{figure}[t]
\begin{center}
\includegraphics*[width=0.94\textwidth,angle=0,clip]{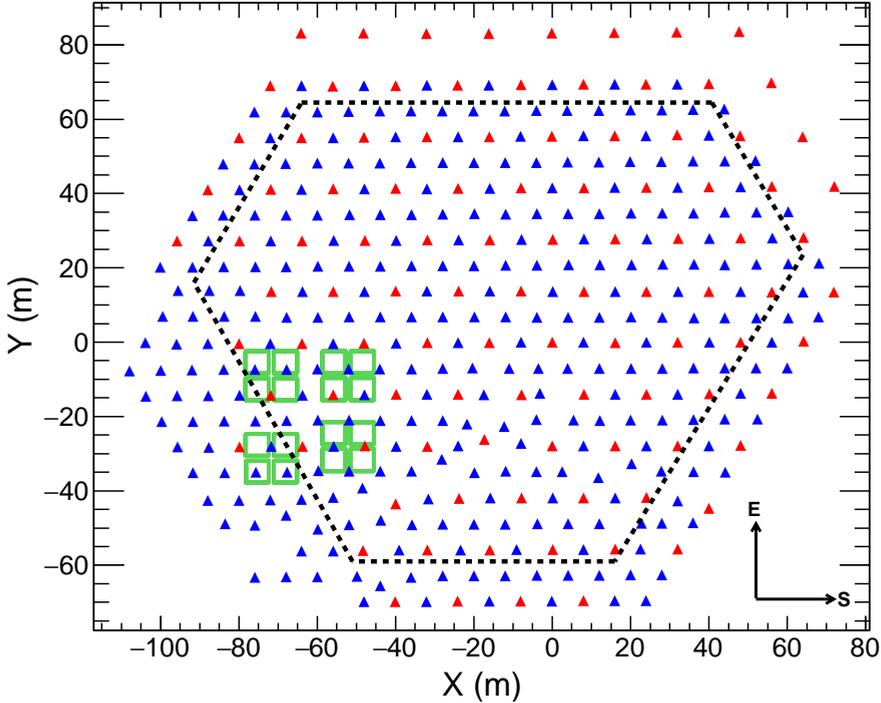}
\caption{Schematic of GRAPES-3 air shower array shows placement of single-PMT
scintillators (\textcolor[rgb]{0,0,1}{$\blacktriangle$}), double-PMT
scintillators (\textcolor[rgb]{1,0,0}{$\blacktriangle$}), and muon telescope
modules (\textcolor[rgb]{0,0.6,0}{$\square$}). The area marked by dotted lines
represents fiducial area under which the showers are selected for analysis.}
\label{fig01}
\end{center}
\end{figure}

The GRAPES-3, which stands for \textbf{G}amma \textbf{R}ay \textbf{A}stronomy at
\textbf{P}eV \textbf{E}nergie\textbf{S} -- phase \textbf{3} is a ground-based
EAS experiment located at Ooty in southern India (11.4$^\circ$N, 76.7$^\circ$E,
and 2200\,m above msl). It is designed to study high energy particles in
different astrophysical settings including acceleration in the atmosphere during
thunderstorms, solar phenomena, energy spectrum and composition of PCRs, and
diffuse $\gamma$-rays. It consists of two major detector components. One of them
comprises an array of 400 scintillator detectors covering a physical area of
25000\,m$^2$ \cite{Gupta05}. The scintillator detectors with an area of 1\,m$^2$
each are placed in a hexagonal geometry with an inter-detector separation of
8\,m as shown in Fig-\ref{fig01}. An additional photomultiplier tube (PMT) is
used in 105 plastic scintillation detectors to overcome the PMT saturation
problem due to large particle densities especially when the shower core lands
close to the detector. Thus, the two-PMT configuration allows the extension of
particle density measurements to over 10,000\,m$^{-2}$ resulting precise
estimation of shower size ($N_e$) for large EAS \cite{Anuj15}.  

The second major component comprises a large area tracking muon telescope (G3MT)
\cite{Hayashi05}. The G3MT consists of 3712 proportional counters (PRCs)
arranged in 16 muon telescope modules with a total area of 560\,m$^2$. The PRCs
are made up of mild steel tube of 600\,cm in length and 10\,cm$\times$10\,cm in
cross-section with wall thickness of 2.3\,mm. Each muon telescope module houses
four layers of PRCs with alternate layers arranged orthogonal to each other.
The layers are sandwiched by 15\,cm thick concrete blocks. Above the muon
modules, 2\,m thick concrete blocks are stacked in the form of an inverted
pyramidal shape that serves as absorber. The entire mass overburden of
$\sim$550\,g$\cdot$cm$^{-2}$ provides 1\,sec($\theta$)\,GeV threshold for muons
incident at zenith angle $\theta$. The four layer configuration allows the
incident muons to be reconstructed into 169 directions covering 2.3\,sr in the
sky with 4$^{\circ}$ accuracy. The GRAPES-3 records $\sim$3$\times$10$^6$ EAS
per day in the energy range of 1\,TeV--10\,PeV and $\sim$4$\times$10$^9$ muons
above 1 GeV.  The muon flux recorded by G3MT is mostly produced by PCRs of
energy 10\,GeV--10\,TeV. The recorded muon flux has been successfully corrected
for atmospheric effects and detector efficiency variations that can be used to
study long-term CR variation and transient phenomenon
\cite{Mohanty2016_1,Arunbabu2017}. The high quality data of G3MT allowed us to
probe interesting physics phenomenon such as measurement of 1.3\,GV electric
potential in thunderclouds \cite{Hari19} and discovery of a muon burst caused by
a transient weakening of Earth's magnetic shield
\cite{Mohanty2016_2,Mohanty2018}.

\section{EAS reconstruction}

\begin{figure}[t]
\begin{center}
\includegraphics*[width=0.97\textwidth,angle=0,clip]{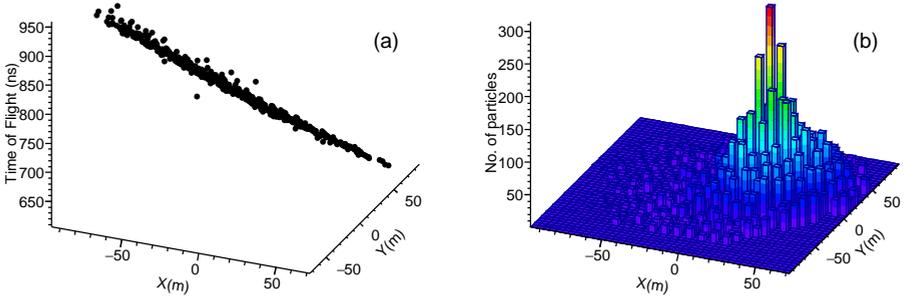}
\caption{A sample EAS (Event time: 20140829-00:11:55.3700155\,IST) recorded at
GRAPES-3 with 365 triggered detectors showing (a) TOF of particles in the shower
disc, and (b) lateral density profile of secondary particles. The shower
parameters are estimated to be $\theta$=37.3$^\circ$, $\phi$=61.3$^\circ$,
$X_c$=22.3\,m, $Y_c$=5.5\,m, $N_e$=1.2$\times$10$^6$, and $s$=1.4.}
\label{fig02}
\end{center}
\end{figure}

The GRAPES-3 experiment records the particle densities and their relative
arrival times in an EAS from all 400 plastic scintillation detectors by using
high precision electronics.  The two-level trigger system is used to eliminate
small but locally developed showers and also very large showers whose cores are
landed very far away. An EAS trigger is generated on the basis of the following
criteria: (i) consecutive three-line coincidence in north-south oriented line of
detectors, which is called Level-0 and (ii) minimum 10 detector hits within the
trigger region called Level-1 \cite{Gupta05}. The energy deposited by the EAS
particles is measured by using charge integrating analog-to-digital converter
(ADC). The ADC information is further converted into equivalent number of muons
by using single muon calibration called as particle densities. The distribution
of particle densities for an EAS is shown in Fig-\ref{fig02}. The relative
arrival time of particles are measured by using a 32-channel high performance
time-to-digital converter (HPTDC) developed in-house \cite{Gupta2012}. The
arrival time distribution for the same EAS is shown in Fig-\ref{fig02}. The true
direction of incident PCR is obtained by fitting the EAS front
\cite{Jhansi2019}.  The observed particle densities are fitted with a lateral
density distribution function known as Nishimura-Kamata-Greisen (NKG) \cite{nkg}
given by Eq-\ref{eqn01} and through a minimization of negative log-likelihood
algorithm using MINUIT. Various EAS parameters including the core location
($X_c$, $Y_c$), shower size ($N_e$), and age ($s$) are obtained from this fit
\cite{Tanaka2012}.

\begin{equation}
    \rho(r_i) = \frac{N_e}{2\pi{r_M}^2}
    \frac{\Gamma(4.5-s)}{\Gamma(s)\Gamma(4.5-2s)}
    \left(\frac{r_i}{r_M}\right)^{s-2} 
    \left(1+\frac{r_i}{r_M} \right)^{s-4.5}
    \label{eqn01}
\end{equation}

\begin{equation}
    r_i = \sqrt{(X_c - X_i)^2 + (Y_c - Y_i)^2}
    \label{eqn02}
\end{equation}

\noindent Here $r_i$ is the distance of $i^{th}$ detector from the shower core
($X_c$, $Y_c$).  $r_M$ is the Moli\`ere radius (distance from the shower core
within which 90\% of the EAS energy is deposited), $r_M$=103\,m for the Ooty
observational level.

\section{Monte Carlo simulations}

The EAS simulations are carried out using CORSIKA which is a widely
used Monte Carlo package for studying the development of EAS in the Earth's
atmosphere \cite{corsika}. The CORSIKA allows the simulation of various
primaries in the entire span of cosmic ray energy spectrum.  It has been
interfaced with several hadronic interaction models such as EPOS-LHC
\cite{eposlhc}, QGSJET01C \cite{qgsjet}, QGSJETII-04 \cite{qgsII}, SIBYLL
\cite{sibyll_2.1}, VENUS \cite{venus}, DPMJET \cite{dpmjet}, and NEXUS
\cite{nexus} for high energy (calculation of cross-sections above 80\,GeV) and
GHEISHA \cite{gheisha}, FLUKA \cite{fluka}, and UrQMD \cite{urqmd} for low
energy interactions. It carries out four dimensional simulations to study the
shower development including various hadronic and electromagnetic interactions,
and decays. The secondary particles are tracked down to the ground level until
they decay or till the kinetic energy is above the user-defined energy
threshold. The physical quantities like position, momentum, and arrival time of
secondary particles can be recorded up to maximum of ten desired observational
levels. The CORSIKA generated secondary particles can be converted into
observables of an experiment by using Geant4, which is a detector simulation
toolkit developed by CERN that allows studying response of various type of
particles and interactions in the material of any arbitrary geometry
\cite{geant}. By this method the simulated data is prepared, which can be
directly compared with the experimental data. 

\begin{figure}[t]
\begin{center}
\includegraphics*[width=0.97\textwidth,angle=0,clip]{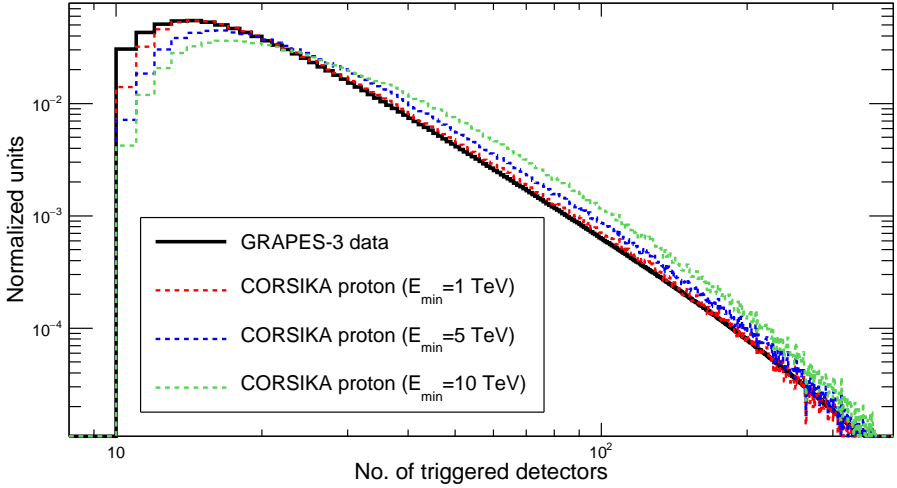}
\caption{Distribution of triggered detectors for GRAPES-3 EAS (solid line), and
Monte Carlo simulations (dotted lines). Showers are selected for cores within
the fiducial area as shown in Fig-\ref{fig01}. All the distributions are
normalized by integral count for
comparison.}
\label{fig03}
\end{center}
\end{figure}

In GRAPES-3, the simulations are performed in three stages: (i) CORSIKA showers
are simulated with a set of user-defined inputs, (ii) the simulated secondary
particles from each shower are subjected to in-house simulation code to do
pseudo random arrival of EAS to generate triggers and also Geant4 response of
collected particles in each plastic scintillation detectors, and (iii)
reconstruction of angle and NKG parameters using TOF and particle densities
respectively. The primary energy for simulation is randomly selected from a
user-defined energy spectrum of $E_{min}$ and $E_{max}$ following a power-law
with a spectral index of -2.7.  The CORSIKA simulated showers are randomly
tossed into the fiducial area (14560\,m$^2$, i.e. 56\% of total physical area)
defined in the GRAPES-3 array shown in Fig-\ref{fig01} to ensure maximum
sampling of spatial distribution. Then the particles which fall over 1\,m$^2$
area of each scintillator are converted into particle density by using
pre-simulated Geant4 calibration. The Geant4 response of the plastic
scintillation detectors with full geometrical implementations are studied
in-depth for gamma, electron, muon, proton, neutron, and pion, that attribute
to most of the secondaries found at the detector level. Each of these particles
is simulated for its response on a wide range of energy and incident direction.
The simulated response of each type of particles is transformed into integral
probability that is further used to calculate the energy deposit for an incident
particle with the inclusion of Poisson fluctuations. Similarly, the single muon
response is also simulated using Geant4. Then the particle density of each
scintillation detector is calculated by using the energy deposited by the
particles, which are passing through the sensitive area of the detector.
Similarly, the first arrival time of the particles for each detector is
accounted as it is directly available in CORSIKA output. Each detector should
have a minimum energy deposit that can pass through discriminator level to be
treated as a triggered detector.  The discriminator level used for scintillators
in the GRAPES-3 experiment is equivalent to 50\% of energy deposit by single
muon response obtained from calibration.  Subsequently, the logical EAS trigger
generation is done as discussed in previous section. The estimated particle
density and TOF are used to calculate the primary properties that may
be used to assess systematics of the measurements.

The simulation parameters can be optimized to reproduce the data. For example,
one such important observable is shown in Fig-\ref{fig03} contains distribution
of triggered detectors from EAS collected by the GRAPES-3. The triggered
detectors distribution for data is generated with $\sim$6.6$\times$10$^8$ EAS
collected during year 2014. In order to reproduce this distribution, the Monte
Carlo simulations are carried out for proton primaries in the angular range of
0--45$^{\circ}$ and 0--360$^{\circ}$ for zenith ($\theta$), and azimuth ($\phi$)
angles respectively.  For this study, SIBYLL and FLUKA combination of hadronic
interaction models are used. Electromagnetic processes are treated by EGS code.
The energy thresholds of secondaries are set to 50\,MeV, 10\,MeV, 1\,MeV, and
1\,MeV for hadrons, muons, electrons, and electromagnetic components
respectively.  The secondary particles are tracked down to Ooty observational
level.  In Fig-\ref{fig03}, the simulated distribution of triggered detectors
for different $E_{min}$ and fixed E$_{max}$=3\,PeV generated with spectral index
of -2.7 are shown.  Here, the distributions are normalized to integral count for
comparison. 

At first, the simulations are carried out for $E_{min}$=10\,TeV to obtain
distribution of triggered detectors.  One can notice the simulated distribution
for 10\,TeV--3\,PeV has large discrepancy in most part of the distribution.  At
the peak position of the data (i.e. $\sim$14 triggered detectors), the
discrepancy with the simulation is found to be $\sim$40\%. The discrepancy
increased to $\sim$90\% at $\sim$100 detectors, reduced to $\sim$66\% at
$\sim$200 detectors, and further dropped to $\sim$50\% at $\sim$300 detectors
and beyond. Especially at smaller number of triggered detectors, the
disagreement indicates the triggering of low energy EAS. This could also be the
cause of discrepancy found at larger side due to relative abundance between low
to high energy EAS which are actually triggered. Motivated by this idea, the
simulations are carried out by reducing $E_{min}$ systematically starting from
5\,TeV, 3\,TeV, and 1\,TeV.  As the simulation energy is lowered the
disagreement, which is seen with $E_{min}$=10\,TeV has reduced from $\sim$40\%
to $\sim$10\% at data peak for $E_{min}$=1\,TeV.  The discrepancies at other
regions are also reduced from $\sim$90\% to $\sim$14\% and from $\sim$66\% to
$\sim$4\% at $\sim$100 and $\sim$200 triggered detectors respectively. Above
$\sim$300 detectors, the spectra of data and 1\,TeV simulation are
indistinguishable. However, the statistical errors in simulation are large
compared to data due to smaller number of simulated high energy EAS. Notably,
the peak position of data and 1\,TeV simulation (i.e. at $\sim$14 triggered
detectors) are matching compared to 10\,TeV.  Also, at the peak of triggered
detectors, there is no disagreement between data and simulation at 1\,TeV
compared to $\sim$34\% found for 10\,TeV.  This confirms the detection of TeV
showers in GRAPES-3. The simulated distributions shown in Fig-\ref{fig03} are
only for $E_{min}$ starting with 1\,TeV, 5\,TeV, and 10\,TeV for better
visibility. The GRAPES-3 EAS are selected for shower cores within the fiducial
area and zenith angle up to 45$^{\circ}$ as done for simulations.

\begin{figure}[t]
\begin{center}
\includegraphics*[width=0.97\textwidth,angle=0,clip]{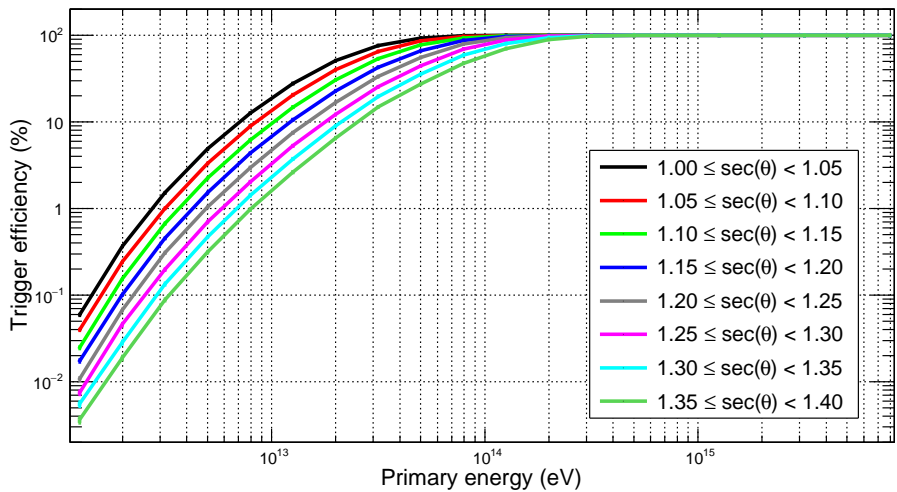}
\caption{Trigger efficiency ($\varepsilon_{tri}$) was estimated by using showers
in each energy bin of 1\,TeV--10\,PeV for various sec($\theta$) intervals.
Shower cores are selected within the fiducial area as shown in Fig-\ref{fig01}.}
\label{fig04}
\end{center}
\end{figure}

\begin{figure}[t]
\begin{center}
\includegraphics*[width=0.97\textwidth,angle=0,clip]{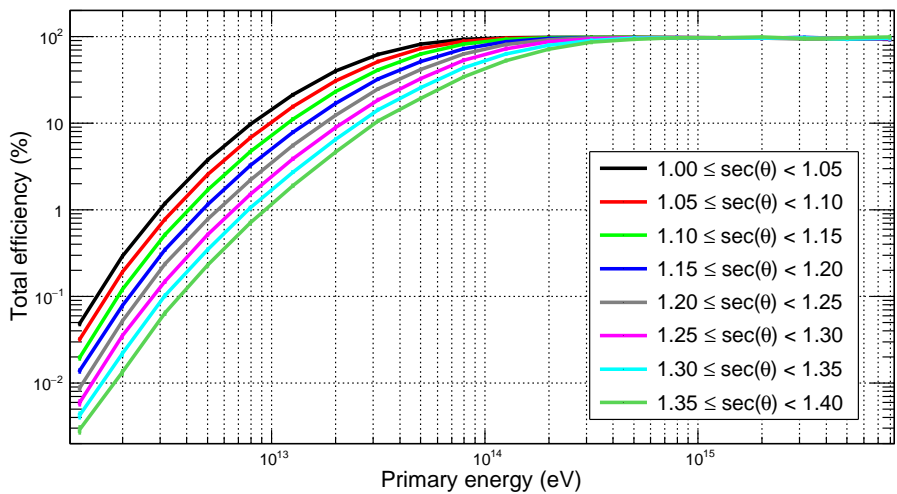}
\caption{Total efficiency ($\varepsilon_{tot}$) was estimated by using
successful NKG reconstruction of triggered showers in each energy bin of
1\,TeV--10\,PeV for various sec($\theta$) intervals.  Shower cores are selected
within the fiducial area as shown in Fig-\ref{fig01}.}
\label{fig05}
\end{center}
\end{figure}

\section{Calculation of efficiencies and acceptance}

\subsection{Trigger ($\varepsilon_{tri}$) and total efficiencies
($\varepsilon_{tot}$)}
By using the optimized $E_{min}$=1\,TeV, the energy range of 1\,TeV--10\,PeV is
used to carry out detailed simulations for further studies. This energy range is
divided into 20 equal intervals in logarithmic scale (width of 0.2). All the
remaining other inputs are kept unchanged.  The proton primaries are simulated
and subjected to trigger generation in each energy interval. The number of
simulated showers in each logarithmic interval is shown in Fig-\ref{tab01}. It
is to be noted that the simulated showers are reused ten times by randomizing
the shower core and secondary particles locations to enhance the statistics. Due
to large area and randomized locations, reuse of showers help to boost the
statistics unbiased. 

The showers are divided into eight equal sec($\theta$) intervals (width of 0.05)
up to $\theta$$\leq$45$^{\circ}$ for further study of angular dependence. The
fraction of triggered showers over total incident is treated as the triggering
efficiency ($\varepsilon_{tri}$) as shown in Fig-\ref{fig04} for various
sec($\theta$) intervals as a function of energy. As discussed in previous
section, the triggering efficiency is tiny ($<$0.1\%) at low energies and
increases to maximum beyond $\sim$100\,TeV. The inclined showers undergoes more
interactions due to increased grammage in the atmosphere as they propagate down
to observational level, which results in reduced particle densities and
subsequently results lower triggering efficiencies. Afterwards, the triggered
showers are reconstructed with NKG and the fit parameters are calculated. The
number of usable showers further drops by a small factor due to inefficiency in
reconstruction.  This reconstruction efficiency is combined with the trigger
efficiency and quoted as total efficiency ($\varepsilon_{tot}$) as shown in
Fig-\ref{fig05} for proton initiated EAS as a function of energy and direction.
These efficiencies determine the amount of usable showers available for
analysis. Also, these inefficiencies have to be corrected in the flux
calculations required for composition studies.

\begin{figure}[t]
\begin{center}
\includegraphics*[width=0.97\textwidth,angle=0,clip]{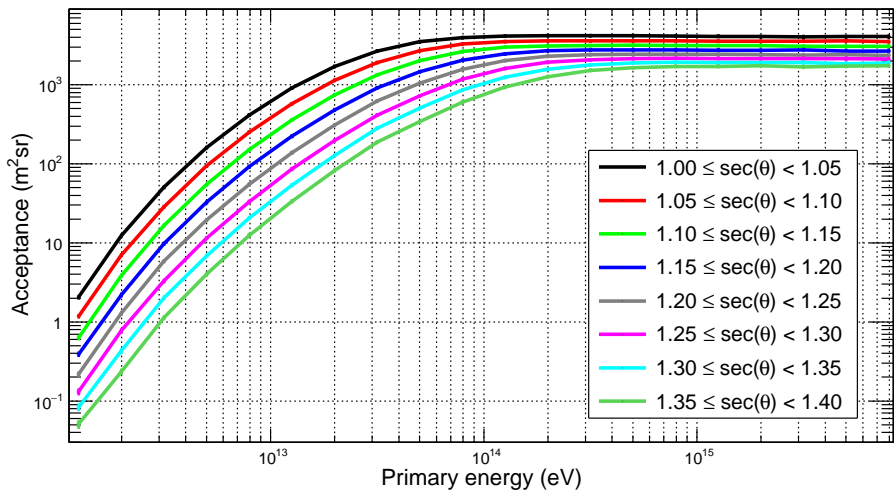}
\caption{GRAPES-3 acceptance for proton initiated showers.}
\label{fig06}
\end{center}
\end{figure}

\subsection{Total acceptance ($\delta_{tot}$)}
The total efficiency ($\varepsilon_{tot}$) calculated by the above mentioned
method is used to determine the acceptance ($\delta_{tot}$) viewed by GRAPES-3
in the sky. The acceptance is defined by the field of view of the detector area
with the inclusion of energy dependent efficiency (Eq-\ref{eqn03}).  The
parameter $\varepsilon_{tot}$ can be treated as a constant for a set of given
energy and zenith angle bin. This can be expressed as Eq-\ref{eqn04} in-terms of
$E$ and $\theta$ to include angular dependent efficiency. 

\begin{equation}
    \delta_{tot}(E) = \int^{\Omega}_{0} 
    A \varepsilon_{tot} (E, \theta) 
    cos \theta d\Omega
    \label{eqn03}
\end{equation}

\begin{equation}
    \delta_{tot}(E) = \frac{\pi A}{2} 
    \sum^{n}_{k=1}
    \varepsilon_{tot} (E, \theta_k)
    (cos 2\theta_k - cos 2\theta_{k+1})
    \label{eqn04}
\end{equation}

\noindent where $A$ = total fiducial area covered (14560\,m$^2$), $n$ = number
of angular intervals, $\theta_{k}$ and $\theta_{k+1}$ = lower and upper edge of
the angular interval, and $E$ = median energy in the logarithmic energy
interval.

Eq-\ref{eqn04} is evaluated for median value of every energy and lower, and
upper limits of angular intervals. Fig-\ref{fig06} shows the total acceptance
for proton primaries measured by GRAPES-3. Also, the total acceptance values are
shown in Table-\ref{tab01} that varies from $\sim$5\,m$^2$sr to
$\sim$21000\,m$^2$sr. Especially at low energies, the total acceptance is larger
compared to direct measurements by satellites and balloon borne experiments that
are $\leq$0.5\,m$^2$sr \cite{dampe,bess,pamela_2,ams2_2,cream_2}.

\section{Discussions}

\afterpage{
\begin{longtable}[tc!]{|R{0.5cm}|R{1.6cm}|R{1.6cm}|R{1.6cm}|C{2.0cm}|R{1.6cm}|}
\hline 
Bin & \multicolumn{3}{c|}{Primary energy (TeV)} & No. of simulated EAS & $\delta_{tot}$ (m$^2$sr) \\
\cline{2-4}
 & \centering E$_{min}$ & \centering E$_{max}$ & \centering E$_{med}$ & & \\ \hline
\endhead
\endfoot
\hline
1   &     1.00 &     1.58 &    1.26 & 5$\times$10$^7$ &     4.72 \\
2   &     1.58 &     2.51 &    1.99 & 3$\times$10$^7$ &    28.10 \\
3   &     2.51 &     3.98 &    3.16 & 2$\times$10$^7$ &   117.72 \\
4   &     3.98 &     6.31 &    5.01 & 1$\times$10$^7$ &   383.11 \\
5   &     6.31 &    10.00 &    7.94 & 5$\times$10$^6$ &  1028.27 \\
6   &    10.00 &    15.85 &   12.59 & 3$\times$10$^6$ &  2380.22 \\
7   &    15.85 &    25.12 &   19.95 & 2$\times$10$^6$ &  4781.71 \\
8   &    25.12 &    39.81 &   31.62 & 1$\times$10$^6$ &  8331.12 \\
9   &    39.81 &    63.10 &   50.12 & 5$\times$10$^5$ & 12326.87 \\
10  &    63.10 &   100.00 &   79.43 & 3$\times$10$^5$ & 16139.69 \\
11  &   100.00 &   158.49 &  125.89 & 2$\times$10$^5$ & 18930.69 \\
12  &   158.49 &   251.19 &  199.53 & 1$\times$10$^5$ & 20632.75 \\
13  &   251.19 &   398.11 &  316.23 & 5$\times$10$^4$ & 21454.41 \\
14  &   398.11 &   630.96 &  501.19 & 3$\times$10$^4$ & 21793.33 \\
15  &   630.96 &  1000.00 &  794.33 & 2$\times$10$^4$ & 21843.11 \\
16  &  1000.00 &  1584.89 & 1258.93 & 1$\times$10$^4$ & 21741.89 \\
17  &  1584.89 &  2511.89 & 1995.26 & 5$\times$10$^3$ & 21656.99 \\
18  &  2511.89 &  3981.07 & 3162.28 & 3$\times$10$^3$ & 21576.61 \\
19  &  3981.07 &  6309.57 & 5011.87 & 2$\times$10$^3$ & 21504.99 \\
20  &  6309.57 & 10000.00 & 7943.28 & 1$\times$10$^3$ & 21403.82 \\
\hline
\caption{Total acceptance ($\delta_{tot}$) of GRAPES-3 for proton initiated
showers.}
\label{tab01}
\end{longtable}
}

Detailed Monte Carlo simulations are carried out by using CORSIKA, Geant4, and
in-house simulation framework to estimate the energy threshold, efficiencies,
and acceptance of proton initiated EAS for GRAPES-3. The initial simulations are
carried out to match the number of triggered detectors distribution obtained
from the data. The simulation energy range is optimized by systematically
reducing lower boundary of the energy range starting from E$_{min}$=10\,TeV,
5\,TeV, 3\,TeV, and 1\,TeV to E$_{max}$=3\,PeV. A reasonably good agreement with
data can be seen by simulations starting from 1\,TeV compared to 10\,TeV as
shown in Fig-\ref{fig03}, which indicates the detection of TeV proton EAS by
GRAPES-3. Further simulations are carried out in the energy range of
1\,TeV--10\,PeV by equally dividing the energy range into logarithmic intervals.
Also, each energy bin is further divided into sec($\theta$) angular intervals
and their dependences are studied. As shown in Fig-\ref{fig04}, the trigger
efficiency at 1\,TeV is $<$0.1\% and reaching to 100\% at $\sim$100\,TeV for
near vertical protons. The trigger efficiency decreases with increasing angle.
By combining this with NKG reconstruction, the total efficiency is estimated as
shown in Fig-\ref{fig05}. Due to the limited information recorded, the
inefficiency of NKG reconstruction reduces the total efficiency by a small
factor in most of the low energy regions. The efficiencies remain maximum at
higher energies. The parameter $\varepsilon_{tot}$ gives the amount of usable
EAS recorded by the GRAPES-3 in each energy interval.  However, at higher
energies, the showers are successfully reconstructed and $\varepsilon_{tri}$ and
$\varepsilon_{tot}$ are unaffected.  Because of the power-law nature of PCR
energy spectrum, majority of the incident PCR are low energy primaries (i.e. in
the given energy range of 1\,TeV--10\,PeV, almost $\sim$98\% are below 10\,TeV).
At low energies, though the efficiencies are tiny, because of the large number
of incident PCR, the GRAPES-3 records most of the low energy primaries useful
for physics analysis.

This can be quantified with the parameter $\delta_{tot}$ for a given experiment
derived by combining $\varepsilon_{tot}$, area, and solid angle as a function of
energy. The geometrical acceptance is estimated to be $\sim$5\,m$^2$sr for
protons of energy starting from 1\,TeV and increasing to $>$20000\,m$^2$sr above
100\,TeV as shown in Table-\ref{tab01}. The minor variations seen in the higher
energy bins are an artifact of statistical fluctuations due to smaller number of
simulated  protons. It is important to note that the geometrical acceptance of
GRAPES-3 has practical implications in measuring sizable amount of low energy
primaries. The large sensitive area ($\sim$2\% of total physical area
25000\,m$^2$) of GRAPES-3 experiment allows to measure proton initiated EAS
energies as small as 1\,TeV. Especially, the acceptance of $\sim$5\,m$^2$sr at
lower energy region is larger than direct measurements from balloon borne
experiments and space probes, which are not more than $\sim$0.5\,m$^2$sr. This
unique capability of GRAPES-3 experiment has been exploited in one of the recent
work where PCR energy spectrum is reported to have fine features in the low
energy region \cite{Fahim19}. A knee-like structure is observed at
(45.4$\pm$0.3)\,TeV.

\section{Conclusions}

The energy threshold of the GRAPES-3 EAS array for primary proton is estimated
to be $\sim$1\,TeV through Monte Carlo simulations with reasonably good
agreement with data. Further detailed simulations are carried out to estimate
various efficiencies as a function of energy and direction. These simulations
are carried out in the energy range of 1\,TeV--3\,PeV and zenith range of
0--45$^\circ$ by dividing into equal intervals. The total efficiency is
estimated to be $\sim$0.1\% for 1\,TeV, and reaching maximum at $>$100\,TeV.
Though, the efficiency is tiny at low energies, the number of usable primaries
are large due to immense flux of low energy PCRs. Similarly, the total
acceptance of EAS array is estimated to be $\sim$5\,m$^2$sr for 1\,TeV, and with
a maximum of $\sim$20000\,m$^2$sr. The acceptance of $\sim$5\,m$^2$sr at 1\,TeV
is large compared to direct measurements. Thus, GRAPES-3's measurements allow to
overlap the energy spectrum from direct measurements at low energies. Also, the
extension of energy measurements beyond knee by GRAPES-3 allows to overlap with
indirect measurements at ultra-high energies. Hence, the GRAPES-3 may provide a
unique handle to bridge the PCR energy spectrum from direct and indirect
measurements.

\begin{acknowledgements}
We thank D.B. Arjunan, A.S. Bosco, V. Jeyakumar, S. Kingston, N.K. Lokre,
K. Manjunath, S. Murugapandian, S. Pandurangan, B. Rajesh, K. Ramadass, R. Ravi,
V. Santhoshkumar, S. Sathyaraj, M.S. Shareef, C. Shobana, R. Sureshkumar, and
other colleagues for their help in running and maintenance of the GRAPES-3
experiment.
\end{acknowledgements}


\end{document}